%
%
%
\def\today{\ifcase\month\or January\or February\or March\or April\or May\or
June\or July\or August\or September\or October\or November\or December\fi
\space\number\day, \number\year}
%
%
\newcount\notenumber

\def\note{\global\advance\notenumber by 1 \footnote{$^{\the\notenumber}$}}
%
%
\newif\ifsectionnumbering
\newcount\eqnumber
\def\cleareqnumber{\eqnumber=0}
\def\numbereq{\global\advance\eqnumber by 1
\ifsectionnumbering\eqno(\the\secnumber.\the\eqnumber)\else\eqno
(\the\eqnumber)\fi}
\def\eqalinno{{\global\advance\eqnumber by 1}
\ifsectionnumbering(\the\secnumber.\the\eqnumber)\else(\the\eqnumber)\fi}
\def\name#1{\ifsectionnumbering\xdef#1{\the\secnumber.\the\eqnumber}
\else\xdef#1{\the\eqnumber}\fi}
\def\nosectionnumbering{\sectionnumberingfalse}
\sectionnumberingtrue
%
%
\newcount\refnumber

\immediate\openout1=refs.tex
\immediate\write1{\noexpand\frenchspacing}
\immediate\write1{\parskip=0pt}
\def\ref#1#2{\global\advance\refnumber by 1%
[\the\refnumber]\xdef#1{\the\refnumber}%
\immediate\write1{\noexpand\item{[#1]}#2}}
\def\tie{\noexpand~}

%
%
\font\twelvebf=cmbx10 scaled \magstep1
\newcount\secnumber

\def\newsection#1.{\ifsectionnumbering\cleareqnumber\else\fi%
	\global\advance\secnumber by 1%
	\bigbreak\bigskip\par%
	\line{\twelvebf \the\secnumber. #1.\hfil}\nobreak\medskip\par\noindent}
%
%
%
\def \sqr#1#2{{\vcenter{\vbox{\hrule height.#2pt
	\hbox{\vrule width.#2pt height#1pt \kern#1pt
		\vrule width.#2pt}
		\hrule height.#2pt}}}}

%
%
%
\newdimen\fullhsize
\def\fiddle{\fullhsize=6.5truein \hsize=3.2truein}
\def\fullline{\hbox to\fullhsize}
\def\mkhdline{\vbox to 0pt{\vskip-22.5pt
	\fullline{\vbox to8.5pt{}\the\headline}\vss}\nointerlineskip}
\def\mkftline{\baselineskip=24pt\fullline{\the\footline}}
\let\lr=L \newbox\leftcolumn
\def\twocolumns{\fiddle
	\output={\if L\lr \global\setbox\leftcolumn=\columnbox
		\global\let\lr=R \else \doubleformat \global\let\lr=L\fi
		\ifnum\outputpenalty>-20000 \else\dosupereject\fi}}
\def\doubleformat{\shipout\vbox{\mkhdline
		\fullline{\box\leftcolumn\hfil\columnbox}
		\mkftline} \advancepageno}
\def\columnbox{\leftline{\pagebody}}
\nosectionnumbering
\magnification=1200
\def\pr#1 {Phys. Rev. {\bf D#1\tie }}
\def\pe#1 {Phys. Rev. {\bf #1\tie}}
\def\pre#1 {Phys. Rep. {\bf #1\tie}}
\def\pl#1 {Phys. Lett. {\bf #1B\tie }}
\def\prl#1 {Phys. Rev. Lett. {\bf #1\tie }}
\def\np#1 {Nucl. Phys. {\bf B#1\tie }}
\def\ap#1 {Ann. Phys. (NY) {\bf #1\tie }}
\def\cmp#1 {Commun. Math. Phys. {\bf #1\tie }}
\def\imp#1 {Int. Jour. Mod. Phys. {\bf A#1\tie }}
\def\mpl#1 {Mod. Phys. Lett. {\bf A#1\tie}}
\def\jhep#1 {JHEP {\bf #1\tie}}
\def\nuo#1 {Nuovo Cimento {\bf B#1\tie}}
\def\tie{\noexpand~}

\parskip=15pt plus 4pt minus 3pt
\headline{\ifnum \pageno>1\it\hfil Schwarzschild metric
beyond linear order $\ldots$\else \hfil\fi}
\font\title=cmbx10 scaled\magstep1
\font\tit=cmti10 scaled\magstep1
\footline{\ifnum \pageno>1 \hfil \folio \hfil \else
\hfil\fi}
\raggedbottom


\overfullrule0pt


\rightline{\vbox{\hbox{RU00-03-B}\hbox{hep-th/0007053}}}
\vfill
\centerline{\title RECOVERY OF THE SCHWARZSCHILD METRIC IN THEORIES}
\centerline{\title WITH LOCALIZED GRAVITY BEYOND LINEAR ORDER}
\vfill
{\centerline{\title Ioannis Giannakis${}^{a}$
and Hai-cang Ren${}^{a, b}$ \footnote{$^{\dag}$}
{\rm e-mail: \vtop{\baselineskip12pt
\hbox{giannak@theory.rockefeller.edu, ren@theory.rockefeller.edu,}}}}
}
\medskip
\centerline{$^{(a)}${\tit Physics Department, The Rockefeller
University}}
\centerline{\tit 1230 York Avenue, New York, NY
10021-6399}
\medskip
\centerline{$^{(b)}${\tit Department of Natural Science, Baruch College of CUNY}}
\centerline{\tit New York, NY 10010} 
\vfill
\centerline{\title Abstract}
\bigskip
{\narrower\narrower
We solve the Einstein equations in the Randall-Sundrum framework with
a static, spherically symmetric matter distribution on
the {\it physical brane} and obtain an approximate
expression for the gravitational field outside the source to
the second order in the gravitational coupling. This expression
when confined on the {\it physical brane} coincides
with the standard form of the Schwarzschild metric. Therefore,
the Randall-Sundrum scenario is consistent with the
Mercury precession and other tests of General Relativity.
\par}
\vfill\vfill\break


\newsection Introduction.%

Motivated by the hierarchy problems of the fundamental interactions,
Randall and Sundrum have proposed an interesting scenario of extra
non-compact dimensions in which four-dimensional gravity emerges
as a low energy effective theory \ref{\randall}{
L. Randall and R. Sundrum, \prl83 (1999) 4690}. This novel proposal
consists of two basic components. The first one is based
on the assumption that ordinary matter and its gauge interactions
are confined within a four dimensional hypersurface, referred to subsequently
as the {\it physical brane}, embedded in a five-dimensional space
\ref{\dim}{K. Akama, {\it Pregeometry}, in Gauge Theory and
Gravitation, edited by K. Kikkawa, N. Nakanishi and H. Nariai,
Springer Verlag, Berlin (1983) 267,
V. A. Rubakov, and M. E. Shaposhnikov, \pl125 (1983) 136,
M. Visser, \pl159 (1985) 22, N. Arkani-Hamed, S. Dimopoulos and
G. Dvali, \pl429 (1998) 263, I. Antoniadis,
N. Arkani-Hamed, S. Dimopoulos and G. Dvali, \pl436 (1998) 257.}.
The second ingredient involves the
realization that a bound state of a five-dimensional graviton
exists and is localized near the {\it physical brane}. Consequently,
gravitational interactions at large distances on the {\it physical
brane} are dominated by the bound state, and thus gravity appears
effectively as four-dimensional. Modifications of this
scenario include supersymmetric generalizations \ref
{\bag}{R. Altendorfer, J. Bagger and D. Nemeschansky, ``{\it Supersymmetric
Randall-Sundrum Scenario}'', hep-th/0003117.}
and models in which gravity
is mediated with metastable graviton bound states \ref{\ruba}{C. Charmousis,
R. Gregory and V. A. Rubakov, ``{\it Wave function of the
radion in a brane world}'', hep-th/9912160; R. Gregory, V. A. Rubakov,
and S. M. Sibiryakov, \prl84 (2000) 5928,
hep-th/0002072; I. Kogan, S. Mouslopoulos, A. Papazoglou,
G. Ross and J. Santiago, ``{\it A three-brane universe: New
phenomenology for the new millenium?}'', hep-th/9912552.}.
Different issues concerning these models have been discussed in
\ref{\por}{C. Csaki, J. Erlich, and T. Hollowood, \pl481 (2000) 107,
hep-th/0003020;
G. Dvali, G. Gabadadze, and M. Porrati, ``{\it 4D Gravity on a
Brane in 5D Minkowski Space''}, hep-th/0005016; \pl484 (2000)
112, hep-th/0002190; R. Gregory,
V. A. Rubakov, and S. M. Sibiryakov, ``{\it Gravity and Antigravity in
a brane world with metastable gravitons}'', hep-th/0003045;
C. Csaki, J. Erlich, T. Hollowood, and J. Terning ``{\it Holographic
RG and Cosmology in Theories with Quasi-Localized Gravity}'',
hep-th/0003076.}.
The Randall-Sundrum model
consists of a three-brane embedded in a five-dimensional space which
is asymptotically anti-de Sitter. A generic form of the metric in this 
space is
$$
ds^2=e^{-2\kappa|y|}{\bar g}_{\mu\nu}dx^\mu dx^\nu+dy^2,
\numbereq\name{\eqena}
$$
the four-dimensional metric ${\bar g}_{\mu\nu}$ being asymptotically flat. 
The asymptotic metric with ${\bar g}_{\mu\nu}
=\eta_{\mu\nu}$ satisfies the vacuum 
Einstein equations
$$
R_{mn}-{1\over 2}Rg_{mn}-\Lambda g_{mn}=0
\numbereq\name{\eqdyo}
$$
with the cosmological constant $\Lambda=-6\kappa^2$. Here and
throughout the 
paper, we adopt the convention that the Greek indices take values 
0-3 and the Latin indices 0-4. The striking feature of this scenario is
the existence of a massless graviton bound state which corresponds  
to the metric fluctuations $h_{\mu\nu}=g_{\mu\nu}-\eta_{\mu\nu}$,
constant in $y$ but nevertheless normalizable because of the 
conformal factor $e^{-2\kappa|y|}$. In the absence of nonzero modes (i.e., for 
a $y$ independent metric $\bar g_{\mu\nu}$) the five-dimensional Einstein 
equations with negative cosmological constant
reduce to the four-dimensional Einstein equations in 
vacuum without a cosmological constant. 
Thus the Randall-Sundrum scenario
can be extended, provided that the metric ${\bar g}_{\mu\nu}$ is Ricci
flat, i.e., corresponds to any
vacuum solution of General Relativity. This can be seen as follows:
The Einstein equations can be rewritten as
$$
R_{\mu\nu}-4\kappa^2g_{\mu\nu}=0, \quad R_{4\mu}=0,
\quad R_{44}-4\kappa^2=0
\numbereq\name{\eqtesse}
$$
with $g_{\mu\nu}=e^{-2\kappa|y|}{\bar g}_{\mu\nu}$.
If we express the five dimensional Ricci tensor in terms of the four dimenional
metric ${\bar g}_{\mu\nu}$, we find
$$
R_{\mu\nu}={\bar R_{\mu\nu}}+4\kappa^2g_{\mu\nu}.
\numbereq\name{\eqfiond}
$$
for a $y$ independent metric ${\bar g}_{\mu\nu}$.
Consequently, ${\bar g}_{\mu\nu}$ obeys ${\bar R_{\mu\nu}}=0$.
An example is the Schwarzschild solution \ref{\haw}{A. Chamblin, S. W. Hawking
and H. S. Reall, \pr61 (2000) 065007.}
$$
ds^2=e^{-2\kappa|y|}\Big[-(1-{{2GM}\over r})dt^2+{1\over {{1-{2GM\over r}}}}dr^2
+r^2d\Omega^2\Big]+dy^2,
\numbereq\name{\eqtria}
$$
the source of which is a mass line extending infinitely in the $y$-direction.
Another example is the pp-wave solution which
describes gravitational waves propagating on the brane
\ref{\gib}{A. Chamblin and G. W. Gibbons, \prl84 (2000) 1090.}.
In the context of a physical situation-a spherically symmetric mass
distribution located on the
{\it physical brane}-nonzero modes are expected to contribute to the dynamics.
The subsequent $y$-dependence of the metric introduces difficulties in
solving the Einstein equations exactly. 
How these modes localize the matter induced curvature near the brane
yet maintain on the brane the asymptotic agreement between the
solutions of the five-dimensional Einstein equations and the
four-dimensional ones has only been shown explicitly for linear gravity.
In this paper, we shall extend such an analysis to the first order of
nonlinearity ( second order in the gravitational coupling )
for the static axially symmetric metric produced by spherically
symmetric matter
that is localized on the brane. 

Similar to the case of linearized gravity, the
solution to the Einstein equations
can be formulated as a solution free from any metric singularity far
away from the source when it is expressed in terms of
coordinates that are straight with respect to the horizon.
We shall eventually transform the solution to a new coordinate
system-natural coordinates of the brane-by
performing a gauge transformation that is
determined by the Israel matching condition \ref{\isr}{W. Israel, \nuo44
(1966) 1.} ( Discussion on these issues concerning linearized gravity
appear in \ref{\are}{I. Y. Arefe'va, M. Ivanov, W. Muck, K. S. Viswanathan,
and I. V. Volovich, ``{\it Consistent linearized gravity in brane
backgrounds}'', hep-th/0004114; M. Ivanov and I. V. Volovich,
{\it Metric fluctuations in brane worlds}, hep-th/9912242; H. Collins
and B. Holdom, {\it Linearized Gravity about a Brane}, hep-th/0006158.}).
We shall work in the weak field approximation and
in a region outside the material source where the Israel
matching condition reduces to the Neumann
boundary condition. \footnote* {The physical brane contributes to
the energy-momentum tensor as well. In the absence of 
matter, the metric $ds^2=e^{-2\kappa |y|}\eta_{\mu\nu}dx^\mu dx^\nu
+dy^2$ implies that 
$T_{\mu\nu}=m{\eta_{\mu\nu}}\delta(y), T_{4\mu}=T_{44}=0$,
where m is a constant.
Its natural extention in the presence of matter is
$T_{\mu\nu}=m\bar g_{\mu\nu}\delta(y), T_{4\mu}=T_{44}=0.$
The linearity of Einstein's equation with respect to the second order 
derivatives leads to Neumann boundary condition. The same condition also 
guarantees the conservation of $T_{\mu\nu}$.}

Our solution, in contrast to the black-string solution [\haw],
will explore both the far away region on the brane and the $AdS_5$ horizon.
The arbitrary parameters of the solution are determined by the Newtonian
limit on the brane.

The apparent gauge dependence of the solution
is the consequence of the incompleteness of the Randall-Sundrum scenario.
Indeed, as it was pointed out in \ref{\vol}{W. Muck, K. S. Viswanathan
and I. V. Volovich, ``{\it Geodesics and Newton's Law in
Brane Backgrounds}'', hep-th/0002132.},
all material particles fall freely towards the $AdS_5$ horizon.
Additional non-gravitational mechanisms are required to stabilize the brane
and confine all material particles within the brane.
In such models, possibly derived from string theory,
the geodesic equation of material particles
would include extra contributions.
But it is still plausible that the Israel matching condition selects a gauge in which 
the motion of material particles within the brane is given by the 
simple four-dimensional geodesic equation 
$$
{d^2x^\mu\over ds^2}+\Gamma_{\rho\lambda}^\mu{dx^\rho\over ds}
{dx^\lambda\over ds}=0.
\numbereq\name{\eqgoun}
$$

In order to describe the real world, the Randall-Sundrum scenario
has to satisfy all the existing tests of General Relativity. Within
the linearized gravity,
it has been shown that  the Newtonian limit as well as the spatial
part of the Schwartzschild metric can be recovered on the
{\it physical brane} for $\kappa r>>1$ with corrections from the
nonzero graviton modes to the $1/r$ terms of the order $O(1/{{\kappa}^2r^3})$
[\randall], \ref{\gar}{J. Garriga and
T. Tanaka, \prl84 (2000) 2778}, [\vol], \ref{\jim}{M. J. Duff and J. T. Liu,
``{\it On the Equivalence of the Maldacena and Randall-Sundrum Pictures}'',
hep-th/0003237; R. Dick and D. Mikulovicz, \pl476 (2000) 363, hep-th/0001013.}.
Thus the scenario implies a deflection angle of light
within the experimental bounds. For the 
precession angle of Mercury, the observation accuracy is up to
the first order of nonlinearity of the metric component $g_{00}$.
Indeed, for 
the weak field expansion of the static spherical metric
$$
ds^2=-(1-{2GM\over r}-\gamma{G^2M^2\over r^2}+...)dt^2
+(1+{2GM\over r}+...)dr^2+r^2d\Omega^2
\numbereq\name{\eqfgih}
$$
the precession angle measured in radians per revolution
is given by \ref{\wei}{S. Weinberg, {\it Gravitation and Cosmology,
Chapter 8}, John Wiley and Sons, Inc. (1972).}
$$
\delta = {{{\pi}GM}\over L}(6+{\gamma})
\numbereq\name{\eqsoinm}
$$
where $L$ is the semilatus rectum of the Mercury orbit and
$\gamma=0$ for the General Relativity.
It is therefore imperative that the large $r$ behavior of $g_{00}$
on the brane in the Randall-Sundrum scenario is examined beyond
linearized approximation. Our 
perturbative solution reproduces the General Relativity result $\gamma=0$.
Thus the Randall-Sundrum
scenario is consistent with all experimental tests of General Relativity.  

For the sake of simplicity, we shall assume that the {\it physical brane} is
located at $y=0$.
In the next section we shall derive the Einstein equations for a static
axially symmetric five dimensional metric, investigate the gauge
symmetries-coordinate trasformations that respect the particular
form of the metric-and discuss the appropriate boundary conditions.
In section three we shall obtain an
exact solution to the linearized equations, a solution which is free of
curvature singularities and describes the gravitational
field produced by a spherically symmetric mass
distribution located on the brane. In section four, we will derive an
approximate expression of the
metric to second order in the gravitational coupling (first order
in non-linearity) which is
valid in a region far from the mass source. This
expression when confined on the {\it physical brane} is in agreement
with the standard four dimensional Schwarzschild metric and
thus reproduces the correct precession angle of Mercury.
Finally we will discuss the physical implications of our results.

\newsection Einstein Equations, Symmetries and Boundary Conditions.%

The most general axially symmetric and static metric in $D=4+1$ 
has the following form
$$
ds^2=e^{-2\kappa|y|}(-e^adt^2+e^bdr^2+e^cr^2d\Omega^2)+dy^2,
\numbereq\name{\eqpente}
$$
where $d\Omega^2=d\theta^2+\sin^2\theta d\phi^2$ is the solid angle on 
$S^2$ and $a$, $b$ and $c$ are functions of $r$ and $y$.
Substituting the metric (\eqpente) into (\eqtesse) we obtain the following
equations 
$$
-a^{\prime\prime}-{2\over r}a^\prime+{1\over 2}a^\prime(-a^\prime+b^\prime
-2c^\prime)-e^{-2\kappa y+b}\Big[\ddot a-5\kappa\dot a-\kappa\dot b-2\kappa
\dot c +{1\over 2}\dot a(\dot a+\dot b+2\dot c)\Big]=0
\numbereq\name{\eqevon}
$$
$$
\eqalign{
a^{\prime\prime}+2c^{\prime\prime}-{2\over r}b^\prime+{4\over r}c^\prime
+{1\over 2}a^\prime(a^\prime&-b^\prime)-c^\prime(b^\prime-c^\prime)\cr
&+e^{-2\kappa y+b}\Big[\ddot b-5\kappa\dot b-\kappa\dot a-2\kappa\dot c+
{1\over 2}\dot b(\dot a+\dot b+2\dot c)\Big]=0\cr}
\numbereq\name{\eqasho}
$$
$$
\eqalign{
c^{\prime\prime}+{4\over r}c^\prime+{{a^\prime-b^\prime}\over r}&+c^{\prime2}
+{1\over 2}(a^\prime-b^\prime)c^\prime\cr
&+e^{-2\kappa y+b}\Big[
\ddot c-\kappa(\dot a+\dot b)-6\kappa \dot c+{1\over 2}\dot c(\dot a+\dot b
+2\dot c)\Big]-2{{e^{b-c}-1}\over {r^2}}=0\cr}
\numbereq\name{\eqhbcv}
$$
$$
\eqalign{
&\ddot a+\ddot b+2\ddot c-2\kappa(\dot a+\dot b+2\dot c)+{1\over 2}(\dot a^2
+\dot b^2+2\dot c^2)=0\cr
&\dot a^\prime+2\dot c^\prime-{2\over r}(\dot b-\dot c)+{1\over 2}a^\prime
(\dot a-\dot b)-c^\prime(\dot b-\dot c)=0\cr}
\numbereq\name{\eqefta}
$$
where the prime denotes a partial derivative with respect to $r$ and the 
dot denotes a partial derivative with respect to $y$. These equations apply 
to the positive side of the brane, i.e., $y>0$, the corresponding
equations to the negative side of the brane, $y<0$, is obtained
by switching the sign of $\kappa$.

For $y$-independent $a$, $b$ and $c$, equations (\eqefta)
are satisfied automatically and equations (\eqevon), (\eqasho)
and (\eqhbcv)
become those of a static, spherically symmetric and Ricci flat metric
albeit not asymptotically anti-de 
Sitter.

We shall discuss the gauge symmetries of these equations, more
specifically the coordinate transformations that respect the
axially symmetric, static form of the metric.
Let's perform a coordinate transformation (gauge transformation) generated
by $v, u$, functions of $r $and $y$ such that
$r \mapsto r+v(r, y)$ and $y \mapsto y+u(r, y)$ and demand that this
particular coordinate transformation respects the form of the
five dimensional metric (\eqpente). We find that the functions
$v, u$ obey the following relations
$$
\eqalign{
&2{\dot v}+{\dot v}^2+e^{-2{\kappa} y-2{\kappa}v+b(r+u, y+v)}
{\dot u}^2=0\cr
&{v^{\prime}}(1+{\dot v})+e^{-2{\kappa} y-2{\kappa}v+b(r+u, y+v)}
{\dot u}(1+{u^{\prime}})
=0\cr}
\numbereq\name{\eqreann}
$$
while the components of the metric transform under these
residual coordinate transformations as follows
$$
\eqalign{
&a(r, y) \mapsto a(r+u, r+v)-2{\kappa}v\cr
&b(r, y) \mapsto b(r+u, y+v)-2{\kappa}v+{\ln {[(1+u^{\prime})^2
+e^{2{\kappa}y+2{\kappa}v-b(r+u, y+v)}v^{\prime2}]}}\cr
&c(r, y) \mapsto c(r+u, y+v)-2{\kappa}v+2{\ln {(1+{u\over r})}}.\cr}
\numbereq\name{\eqrbcv}
$$
In what follows, we shall construct the solutions to the Einstein equations in 
two different coordinate systems, or equivalently, in
two different gauges [\gar], \ref{\muy}{Y. Myung, G. Kang, and H. Lee,
\pl478 (2000) 294, hep-th/0001107.}.
The solution in one coordinate
system-the one in which the coordinates are
straight with respect to the horizon-is free from metric
singularities far away from the source. We shall denote this solution by $a(r,y)$, 
$b(r,y)$ and $c(r,y)$ without superscripts and we shall impose the following
conditions 
$$
\lim_{|y|, r\to\infty}a=\lim_{|y|, r\to\infty}b
=\lim_{|y|, r\to\infty}c=0
\numbereq\name{\eqexy}
$$
independent of the order of the limits. The other set of coordinates are the
natural coordinates of the brane. In this coordinate
system the solution is given by the right
hand side of (\eqrbcv) with the parameters $u$ and $v$
chosen appropriately to satisfy the
Israel matching condition. We will denote the solution in this gauge
by $a^P(r,y)$,
$b^P(r,y)$ and $c^P(r,y)$, where the superscript $P$ means ``physical''.
Outside
the material source, they satisfy the Neumann boundary condition,
$\dot a^P(r,0)=\dot b^P(r,0)=\dot c^P(r,0)=0.$
In the subsequent sections, we shall use the following notation
$$
\eqalign{
a^P(r,y)&=a(r,y)+\delta a(r,y)\cr
b^P(r,y)&=b(r,y)+\delta b(r,y)\cr
c^P(r,y)&=c(r,y)+\delta c(r,y)\cr}
\numbereq\name{\eqasnbcl}
$$
where $\delta a$, $\delta b$ and $\delta c$ are not necessarily infinitesimal.

The Newtonian limit is specified by the asymptotic behavior of $g_{00}$ 
on the brane for large $r$, i.e.
$$
a^P(r,y)=1-{2GM\over r}-\cdots
\numbereq\name{\eqalves}
$$
with $G$ being the four-dimensional gravitational constant and $M$ the total mass of 
the sourse. For a weak gravitational field, $GM<<r$, we have the following 
perturbative expansions for the solution that is free of metric singularities
$$
\eqalign{
&a(r,y)=a_{0}(r,y)+a_{1}(r,y)+\cdots, \qquad  b(r,y)=b_{0}(r,y)+b_{1}(r,y)+\cdots \cr
&c(r,y)=c_{0}(r,y)+c_{1}(r,y)+\cdots\cr},
\numbereq\name{\eqoxtw}
$$
and for the physical solution
$$
\eqalign{
&a^P(r,y)=a_{0}^P(r,y)+a_{1}^P(r,y)+\cdots, 
\qquad  b^P(r,y)=b_{0}^P(r,y)+b_{1}^P(r,y)+\cdots \cr
&c^P(r,y)=c_{0}^P(r,y)+c_{1}^P(r,y)+\cdots\cr}.
\numbereq\name{\eqoxtw}
$$
Similarly we can expand the parameters of transformation
$$
\eqalign{
v(r, y)&=v_{0}(r, y)+v_{1}(r, y)+\cdots\cr
u(r, y)&=u_{0}(r, y)+u_{1}(r, y)+\cdots.\cr}
\numbereq\name{\eqacoiuxn}
$$
All quantities with the subscript ``0'' are of the order $GM$ and those with 
the subscript ``1'' are of the order $G^2M^2$. In the next section we shall
derive an exact solution to the first order in the gravitational coupling
and we will discuss its physical significance.

\newsection Linearized Solution.%

For a weak gravitational field,
we obtain the linearized Einstein equations, i.e.
$$
\eqalign{
&-a_{0}^{\prime\prime}-{2\over r}a_{0}^\prime-e^{-2\kappa|y|}
\Big[\ddot a_{0}-5\kappa\dot a_{0}-\kappa\dot b_{0}-2\kappa\dot c_{0} \Big]=0,\cr
&a_{0}^{\prime\prime}+2c_{0}^{\prime\prime}-{2\over r}b_{0}^\prime+
{4\over r}c_{0}^\prime
+e^{-2\kappa y}\Big[\ddot b_{0}-5\kappa\dot b_{0}-\kappa\dot a_{0}
-2\kappa\dot c_{0}\Big]=0,\cr
&c_{0}^{\prime\prime}+{4\over r}c_{0}^\prime
+{{a_{0}^\prime-b_{0}^\prime}\over r}
+e^{-2\kappa y}\Big[\ddot c_{0}-\kappa(\dot a_{0}+\dot b_{0})-6\kappa \dot c_{0}
\Big]-{2\over r^2}(b_{0}-c_{0})=0,\cr
&\ddot a_{0}+\ddot b_{0}+2\ddot c_{0}-2\kappa(\dot a_{0}
+\dot b_{0}+2\dot c_{0})=0\cr
&\dot a_{0}^\prime+2\dot c_{0}^\prime-{2\over r}(\dot b_{0}-\dot c_{0})=0.\cr}
\numbereq\name{\eqenea}
$$

Initially, we shall seek solutions that satisfy equation (\eqexy).
The fourth of these equations is particularly simple, and the only 
solution consistent with the condition (\eqexy) is 
$$
a_{0}+b_{0}+2c_{0}=0.
\numbereq\name{\eqdeka}
$$
Similarly, the last of equations (\eqenea)  can be integrated to yield
$$
a_{0}^\prime+2c_{0}^\prime-{2\over r}(b_{0}-c_{0})=\phi(r)
\numbereq\name{\eqeleu}
$$
where $\phi$ is an arbitrary function of $r$. The boundary condition (\eqexy)
then demands that
$$
a_{0}^\prime+2c_{0}^\prime-{2\over r}(b_{0}-c_{0})=0.
\numbereq\name{\eqmauro}
$$
Note that the linearized Schwarzschild solution [\haw] satisfies (\eqeleu)
but not (\eqmauro) . Substituting (\eqdeka) into the first of equations (\eqenea)
and using the Green's function found in \ref{\giddings}{S. Giddings, E. Katz
and L. Randall, \jhep0003 (2000) 023.}, we obtain
$$
a_{0}(r, y)=-{8GM\over {3{\pi}{\kappa}}}e^{2\kappa y}\int_0^\infty dppj_{0}(pr)
{{{K_2(\hat pe^{\kappa y})}}\over {K_1(\hat p)}}
\numbereq\name{\eqkoulh}
$$
where $\hat p=p/\kappa$. $K_\nu(z)$ is the modified 
Bessel function of the second kind and $j_0(x)$ is the spherical Bessel 
function. The choice of the coefficient of the equation (\eqkoulh) is dictated
by the requirement that it reproduces the correct Newtonian limit.
Combining (\eqdeka) and (\eqmauro), we obtain a first order differential
equation  for $b$
$$
b_{0}^\prime+{3\over r}b_{0}+{a_{0}\over r}=0,
\numbereq\name{\eqamana}
$$
the solution of which reads
$$
b_{0}(r, y)={8GM\over {3{\pi}{\kappa}}}e^{2\kappa y}\int_0^\infty dpp
{{j_1(pr)}\over {pr}}
{{{K_2(\hat pe^{\kappa y})}}\over {K_1(\hat p)}}.
\numbereq\name{\eqanatola}
$$
It follows then from (\eqdeka) then that
$$
c_{0}(r, y)={8GM\over {3{\pi}{\kappa}}}e^{2\kappa y}\int_0^\infty dpp
{1\over 2}[j_0(pr)-{{j_1(pr)}\over {pr}}]
{{K_2(\hat pe^{\kappa y})}\over {K_1(\hat p)}}.
\numbereq\name{\eqze}
$$
It is straightforward to verify that 
the solutions (\eqkoulh), (\eqanatola) and (\eqze) satisfy the
two of the remaining
equations (\eqenea) which we have not used. 

If either $r$ or $\zeta\equiv{1\over\kappa}e^{\kappa y}$ becomes large, i.e, 
$\kappa r>>1$ or $\kappa z>>1$, the integrals (\eqkoulh),
(\eqanatola) and (\eqze) are dominated by the region
where $\hat p<<1$. The modified Bessel function in the
denominator, $K_1(\hat p)\simeq {1\over\hat p}$, and the
integrals can be carried out explicitly. We find that
$$
\eqalign{
a_{0}(r, \zeta)&=-{{4GM}\over 3}{{2r^2+3{\zeta}^2}\over {(r^2
+{\zeta}^2)^{3\over 2}}}=-{{4GM}\over 3s}(2+\cos^2\theta)\cr
b_{0}(r, \zeta)&={{4GM}\over 3}{1\over\sqrt{r^2+\zeta^2}}=
{{4GM}\over 3s}\cr
c_{0}(r, \zeta)&={{2GM}\over 3}{{r^2+2{\zeta}^2}\over {(r^2
+{\zeta}^2)^{3\over 2}}}={{2GM}\over 3s}(1+\cos^2\theta)\cr}
\numbereq\name{\eqaoibcv}
$$
where we have introduced four-dimensional spherical
coordinates $s$ and $\theta$ via $r=s\sin\theta$ and $\zeta=s\cos\theta$.
In terms of $s$, and $\theta$ the physical brane corresponds to
$\cos\theta={1\over {{\kappa}r}}$. We are primarily interested in the
region with $\cos{\theta} << 1$.
The result for $a_0(r,y)$ was obtained in [\giddings]. Here we would like to 
emphasize that the expressions on the right hand sides
of (\eqaoibcv) approximate the entire region of $\kappa s>>1$.
So it does interpolate the large $r$ region on the brane and
the large $z$ region near the horizon. Furthermore, as is shown 
in the appendix, systematic inclusion of the higher order
terms of $K_1(\hat p)$ will lend us a large $s$ expansion of the type
$$
{1\over s}\sum_{n=0}^\infty{P_n(\theta; \ln\kappa s)\over s^{2n}}
\numbereq\name{\eqapnvh}
$$
where $P_n(\theta,\ln\kappa s)$ is a $n$-th order polynomial  
in $\ln\kappa s$ with $\theta$-dependent coefficients. 

It is straightforward to check that while $a_0$ satisfies the Neumann
boundary condition, $b_0$ and $c_0$ fail to do so.
In order to transform the solution to the physical one we need to
determine the parameters $v_0$ and $u_0$.
Equations (\eqreann) read to first order as
$$
{\dot v}_{0}=0, \qquad v^{\prime}_{0}+e^{-2\kappa y}{\dot u}_{0}=0
\numbereq\name{\eqrojv}
$$
while the transformations of the components of the metric become
$$
{\delta}a_{0}=-2{\kappa}v_{0}, \qquad {\delta}b_{0}=-2{\kappa}v_{0}
+2u^{\prime}_{0},
\qquad {\delta}c_{0}=-2{\kappa}v_{0}+2{{u_0}\over r}.
\numbereq\name{\eqrpom}
$$
We can easily verify that the Einstein equations (\eqenea) remain
invariant under the transformations (\eqrojv).
Thus, the physical solution will be given by
$$
a^{P}_{0}=a_{0}+{\delta}a_{0}, \qquad b^{P}_{0}=b_{0}+{\delta}b_{0},
\qquad c^{P}_{0}=c_{0}+{\delta}c_{0}.
\numbereq\name{\eqrvmc}
$$
We impose Neumann boundary condition on $b_{0}^P$
on the brane
${\dot b}_{0}^P|_{y=0}=0$. Under a coordinate transformation
generated by $v_0, u_0$, we find that
$$
{\dot b}_{0}^P|_{y=0}={{-4GM}\over {{\kappa}r^3}}+2u^{\prime}_0=
{{-4GM}\over {{\kappa}r^3}}-2v^{\prime\prime}_0=0
\numbereq\name{\eqasonvc}
$$
Thus,
$$
v_{0}(r)=-{{GM}\over {3{\kappa}r}}, \qquad u_{0}(r, y)=
-{{GM}\over {6{\kappa}^2r^2}}
e^{2\kappa y}+{\chi}(r)
\numbereq\name{\eqrvvz}
$$
where $\chi$ is an arbitrary function of $r$ and is chosen
$\chi=-{{2GM}\over 3}$ in order to recover the
standard Schwarzschild metric on the {\it physical brane}.
The physical solution to first order takes then
the form
$$
\eqalign{
a^{P}_{0}&=-{8GM\over {3{\pi}{\kappa}}}
e^{2\kappa y}\int_0^\infty dppj_0(pr)
{{{K_2(\hat pe^{\kappa y})}}\over {K_1(\hat p)}}+{2GM\over 3r}\cr
b^{P}_{0}&={8GM\over {3{\pi}{\kappa}}}e^{2\kappa y}\int_0^\infty dpp
{{j_1(pr)}\over {pr}}
{{{K_2(\hat pe^{\kappa y})}}\over {K_1(\hat p)}}+{2GM\over 3r}+{{2GM}\over
{3{\kappa^2}r^3}}e^{2{\kappa}y}\cr
c^{P}_{0}&={8GM\over {3{\pi}{\kappa}}}e^{2\kappa y}\int_0^\infty dpp
{1\over 2}[j_0(pr)-{{j_1(pr)}\over {pr}}]
{{K_2(\hat pe^{\kappa y})}\over {K_1(\hat p)}}-{2GM\over 3r}
-{{GM}\over
{3{\kappa^2}r^3}}e^{2{\kappa}y}\cr}
\numbereq\name{\eqrvzu}
$$
It is straightforward to verify that $h_0^P=a_0^P+b_0^P+2c_0^P=0$.
Using equations (\eqaoibcv), the metric on the {\it physical brane}
for  $r>>1/\kappa$ becomes
$$
ds^2=-(1-{2GM\over r}+\cdots)dt^2+
(1+{{2GM}\over r}+\cdots)dr^2+r^2d\Omega^2,
\numbereq\name{\eqtud}
$$
thus reproducing the standard form of the Schwarzschild metric.
The dots in equation (\eqtud) represent terms of order $O({{G^2M^2}
\over r^2})$
and higher.

\newsection Second Order Approximate Solution.%

In this section, we will go beyond the first order solution (linearized)
and find an approximate solution to second order in the
gravitational coupling of the
Einstein equations. Inserting the ansatz (\eqoxtw) to equations
(\eqevon)-(\eqefta) and keeping
terms to second order, we find
 $$
\eqalign{
-a_{1}^{\prime\prime}-{2\over r}a_{1}^\prime+{1\over 2}a_{0}^\prime
(-a_{0}^\prime+b_{0}^\prime
-2c_{0}^\prime)&-e^{-2\kappa y}\Big[\ddot a_{1}-5\kappa\dot a_{1}
-\kappa\dot b_{1}
-2\kappa\dot c_{1} +{1\over 2}\dot a_{0}(\dot a_{0}+\dot b_{0}+2\dot c_{0})\cr
&+b_{0}(\ddot a_{0}-5\kappa\dot a_{0}-\kappa\dot b_{0}
-2\kappa\dot c_{0}) \Big]=0\cr}
\numbereq\name{\eqzete}
$$
$$
\eqalign{
&\ddot a_{1}+\ddot b_{1}+2\ddot c_{1}-2\kappa(\dot a_{1}+\dot b_{1}
+2\dot c_{1})+{1\over 2}(\dot a_{0}^2
+\dot b_{0}^2+2\dot c_{0}^2)=0\cr
&\dot a_{1}^\prime+2\dot c_{1}^\prime-{2\over r}(\dot b_{1}-\dot c_{1})
+{1\over 2}a_{0}^\prime
(\dot a_{0}-\dot b_{0})-c_{0}^\prime(\dot b_{0}-\dot c_{0})=0\cr}
\numbereq\name{\eqgewrg}
$$
If we introduce $h_1=a_1+b_1+2c_1$
it follows then from (\eqgewrg) and taking into account equation
(\eqaoibcv) that
$$
{\dot h}_1={{{\kappa}G^2M^2{\zeta}^2}\over
{(r^2+{\zeta}^2)^2}}\Big [ {{11}\over {18}}+{5\over 9}{{\zeta^2}\over
{(r^2+{\zeta}^2)}}+{3\over 2}{{\zeta^4}\over
{(r^2+{\zeta}^2)^2}}\Big ].
\numbereq\name{\eqaristo}
$$
Consequently, integration over $y$ yields
$$
h_1=-{{G^2M^2}\over
{(r^2+{\zeta}^2)}}\Big [ {{25}\over {36}}+{7\over {18}}{{\zeta^2}\over
{(r^2+{\zeta}^2)}}+{1\over 4}{{\zeta^4}\over
{(r^2+{\zeta}^2)^2}}\Big ].
\numbereq\name{\eqaristofa}
$$
Taking into account (\eqdeka), equation (\eqgewrg) becomes
$$
a_{1}^{\prime\prime}+{2\over r}a_{1}^\prime+e^{-2\kappa y}
\Big[\ddot a_{1}-4\kappa\dot a_{1}\Big]=a_{0}^\prime b_{0}^\prime
-e^{-2\kappa y}b_{0}^\prime(\ddot a_{0}-4\kappa\dot a_{0})+e^{-2\kappa y}
{\kappa}{\dot h}_1
\numbereq\name{\eqgewrgat}
$$
with the right hand side completely known.
We proceed to introduce spherical coordinates $s, \theta$ such that
$r=s{\cos\theta}$, $z={1\over {\kappa}}e^{\kappa y}=s{\sin\theta}$.
In terms of these new variables, equation (\eqgewrg) takes the following form
$$
\Big[{1\over {s^3}}{{\partial}\over {\partial s}}
(s^3{{\partial}\over {\partial s}})+{1\over {s^2{\sin^2{\theta}}}}
{{\partial}\over {\partial\theta}}({\sin^2{\theta}}{{\partial}\over {\partial\theta}})
-{3\over s}{{\partial}\over {\partial s}}
+{3\over {s^2}}{\tan{\theta}}{{\partial}\over {\partial\theta}}\Big]a_{1}
(s, \theta)=f_1+f_2+f_3
\numbereq\name{\eqbcud}
$$
where  
$$
\eqalign{
f_1&=a_{0}^\prime b_{0}^\prime=-{{16G^2M^2}\over 9}
{{(2r^2+5{\zeta}^2)r^2}\over {(r^2+{\zeta}^2)^4}}=-{{16G^2M^2}\over 9}
{{2+{\cos^2{\theta}}-3{\cos^4{\theta}}}\over {s^4}}\cr
f_2&=-e^{-2\kappa y}b_{0}^\prime(\ddot a_{0}-4\kappa\dot a_{0})=
{{16G^2M^2}\over 3}
{{5{\zeta}^4}\over {(r^2+{\zeta}^2)^4}}={{16G^2M^2}\over 3}
{{5{\cos^4{\theta}}}\over {s^4}}\cr
f_3&=e^{-2\kappa y}{\kappa}({\dot a}_{1}+{\dot b}_{1}+2{\dot c}_{1})=
{{16G^2M^2}\over 9}{1\over {s^4}}\Big[ {{11}\over {32}}+
{{5}\over {16}}{\cos^2{\theta}}+{{27}\over {32}}{\cos^4{\theta}} \Big]\cr}
\numbereq\name{\eqaokjg}
$$
Equation (\eqbcud) is a second order inhomogeneous linear
differential equation, whose solution is a superposition of a special solution
and a solution of the corresponding homogeneous
equation.
In order to derive the contribution to the solution from the inhomogeneous
terms $f_i, i=1,2,3$ we make the ansatz $\phi_i=
{{A_i+B_{i}\cos^2{\theta}+C_{i}\cos^4{\theta}}\over {s^2}}, i=1,2,3$
with $A_i, B_i, C_i$ constants. Then by demanding that
$$
\Big[{1\over {s^3}}{{\partial}\over {\partial s}}
(s^3{{\partial}\over {\partial s}})+{1\over {s^2{\sin^2{\theta}}}}
{{\partial}\over {\partial\theta}}({\sin^2{\theta}}{{\partial}\over {\partial\theta}})
-{3\over s}{{\partial}\over {\partial s}}
+{3\over {s^2}}{\tan{\theta}}{{\partial}\over {\partial\theta}}\Big]\phi_i
(s, \theta)=f_i, \quad i=1,2,3
\numbereq\name{\eqbbco}
$$
we calculate the coefficients $A_i, B_i, C_i$ and thus determine the
part of the solution which corresponds to the inhomogeneous terms
$$
\eqalign{
\phi_1&=-{{G^2M^2}\over {s^2}}({8\over 9}
+{4\over 9}{\cos^2{\theta}}+{8\over 9}
{\cos^4{\theta}}), \qquad
 \phi_2=-{40\over 9}{{G^2M^2}\over {s^2}}{\cos^4{\theta}}\cr
\phi_3&={{G^2M^2}\over {s^2}}({7\over {36}}+{5\over {36}}
{\cos^2{\theta}}-{1\over 4}
{\cos^4{\theta}}).\cr}
\numbereq\name{\eqahoz}
$$
Thus, a special solution of the equation takes the form
$$
\phi=\phi_1+\phi_2+\phi_3={{G^2M^2}\over {s^2}}
(-{{25}\over {36}}-{{11}\over {36}}{\cos^2{\theta}}
-{{67}\over {12}}{\cos^4{\theta}}).
\numbereq\name{\equdoviz}
$$
The most general form of the solution to the homogeneous equation, which is 
consistent with the large $\kappa s$ approximation can be put in the form
$${1\over s^2}[F_0(\theta)+F_1(\theta)\ln\kappa s+...+F_m(\theta)\ln^m
\kappa s].$$ 
As is shown in the appendix, the ordinary differential equation satisfied by 
$F_m(\theta)$ can be converted into a hypergeometric equation with the two 
linearly independent solutions
$$w_1(\theta)={\cos^3{\theta}}{\cot{\theta}}$$
and
$$w_2(\theta)=-{1\over 2}-{3\over 4}{\cos^2{\theta}}
+{3\over 8}{\cos^3{\theta}}{\cot{\theta}}
{\ln {{1-\sin{\theta}}\over {1+\sin{\theta}}}}.
\numbereq\name{\eqlembo}.$$
$w_1$ is singular at $\theta=0$ ($AdS_5$ horizon) and it should not
be included while $w_2$ is finite at 
$\theta={\pi\over 2}$, its inclusion though would produce on the
brane terms of the
form ${1\over r^2}{\ln^m{\kappa}r}$, in disagreement
with the standard Schwarzschild metric. Thus we expect
$F_m(\theta)=0$. Similarly we expect that
$F_{m-1}(\theta)$, ..., $F_1(\theta)$ are equal to zero and we are left with 
$a_1=\phi+{\eta}w_2$, where $\eta$ is a constant to be determined.

Similarly to the solution of the linearized equations
the physical solution to second order will be given by
$$
a^{P}_{1}=a_{1}+{\delta}a_{1}, \qquad b^{P}_{1}=b_{1}+{\delta}b_{1},
\qquad c^{P}_{1}=c_{1}+{\delta}c_{1}.
\numbereq\name{\eqsta0hs}
$$
The parameters of transformation that generate the residual coordinate
transformations $v_1, u_1$ obey the constraints
(\eqreann) to second order
$$
\eqalign{
2{\dot v_{1}}&=-e^{-2\kappa y}{\dot u}_{0}^{2}=-e^{2\kappa y}
v_{0}^{\prime 2}\cr
{v_{1}^{\prime}}+e^{-2\kappa y}{\dot u_{1}}&=-e^{-2\kappa y}
(b_{0}-2{\kappa}v_{0}){\dot u_{0}}-e^{-2\kappa y}{\dot u_{0}}{u_{0}^{\prime}}\cr}
\numbereq\name{\eqrealm}
$$
where we have used equation (\eqrojv),
while the components of the metric transform to second order as
$$
\eqalign{
&{\delta}a_1=a_{0}^{\prime}u_{0}+{\dot a}_{0}v_0-2{\kappa}v_1\cr
&{\delta}b_1=b_{0}^{\prime}u_{0}+{\dot b}_{0}v_{0}-2{\kappa}v_1
+2u_{1}^{\prime}-u_{0}^{\prime2}
+e^{2\kappa y}v_{0}^{\prime2}\cr
&{\delta}c_1=c_{0}^{\prime}u_{0}+{\dot c}_{0}v_{0}-2{\kappa}v_{1}
+2{{u_1}\over r}
-{{u_{0}^{2}}\over r^2}\cr}
\numbereq\name{\eqgewrgiou}
$$
and
$$
\delta h_1 \equiv {\delta}a_1+{\delta}b_1+2{\delta}c_1=
-8{\kappa}v_1+2u_1^{\prime}+{4\over r}u_1-u_0^{\prime 2}+
e^{2\kappa y}v_0^{\prime 2}-{2u_0^2\over r^2}.
\numbereq\name{\eqaristra}
$$
The constraints (\eqrealm) yield the following expressions for
$v_1$ and ${\dot u_{1}}$,
$$
\eqalign{
{\dot v_{1}}&=-{1\over {2 \kappa}}e^{2\kappa y}v_{0}^{\prime 2}+{\lambda}(r)=
-e^{2\kappa y}{{G^2M^2}\over {36{\kappa}^3r^4}}+{\lambda}(r)\cr
{\dot u_{1}}&=e^{2\kappa y}
(b_{0}-2{\kappa}v_{0}+{u_{0}^{\prime}})v_{0}^{\prime}
-e^{2\kappa y}{v_{1}^{\prime}}\cr}
\numbereq\name{\eqhaniwta}
$$
where $\lambda(r)$ is an arbitrary fucntion of $r$ which can be determined
by enforcing the brane bending condition-Neumann boundary condition
to second order: more specifically,
$$
({\dot h}_1+{\delta}{\dot h}_{1})|_{y=0}=2(-{\lambda}^{\prime\prime}-
{2\over r}{\lambda}^{\prime}-{{17G^2M^2}\over {36{\kappa}r^4}})=0
\numbereq\name{\eqkatergia}
$$
We find that ${\lambda}(r)=-{{17G^2M^2}\over {72{\kappa}r^2}}$ and
consequently that
$$
v_{1}(r, y)=-e^{2\kappa y}{{G^2M^2}\over {36{\kappa}^3r^4}}
-{{17G^2M^2}\over {72{\kappa}r^2}}
\numbereq\name{\eqdilberis}
$$
A comment regarding the way Neumann boundary condition is met
appears necessary at this point.
Unlike the linearized solutions, which we have the exact expressions
at our disposal, here 
only the approximate expressions at large $\kappa s$ are available. So we can 
not satisfy the Neumann boundary condition perfectly. On the other hand, it 
is easy to show that for large $\kappa r$ on the brane, ${\dot a}_1$, 
${\dot b}_1$ and ${\dot c}_1$ are dominated by terms of order
${G^2M^2\over \kappa r^4}$. It is this 
order that has to be cancelled by the gauge transformation in conformity 
to the Neumann boundary condition. In another word, ${\dot a}^P$,
${\dot b}^P$ and 
${\dot c}^P$ should be of higher order
than ${G^2M^2\over \kappa r^4}$ at $y=0$
and should vanish exactly when higher powers of $1/\kappa s$ are restored 
in the expressions of $a^P$, $b^P$ and $c^P$.  

Using equations (\eqsta0hs), (\eqgewrgiou) and (\eqdilberis) we
derive the following expression for $a^P_{1}$
$$
\eqalign{
a^P_{1}=-{{G^2M^2}\over {s^2}}&\Big [{{25}\over {36}}
+{{16}\over 9}{\csc\theta}-{{17}\over {36}}{\csc^2{\theta}}+(
{{11}\over {36}}+{4\over 3}{\csc\theta}-{1\over {18}}{\csc^4{\theta}})
{\cos^2{\theta}}\cr
&+({{67}\over {12}}-{2\over 3}{\csc\theta}){\cos^4{\theta}}+{\eta}w_2(\theta)
\Big ] \cr}
\numbereq\name{\eqtzanos}
$$
It is straightforward to verify that
$a^P_{1}$ satisfies Neumann boundary condition
${\dot a}^P_{1}(r, 0)=0$ only if
$\eta=0$. On the {\it physical brane} and for
$\kappa r>>1$, the time component of the metric to second order reads
$$
-g_{00}=e^{a_{0}^P+a_{1}^P}=1-{{2GM}\over r}+O({{G^3M^3}\over r^3})
\numbereq\name{\eqgiannou}
$$
which is consistent with the General Relativity prediction $\gamma=0$.

For completeness we will determine the expression for
$b_1$ and $c_1$ on the brane and for $\kappa r>>1$.
Using the fact that $h_1=a_1+b_1+2c_1$, the second of equations (\eqgewrg)
takes the form
$$
{\dot b}_1^{\prime}+{3\over r}{\dot b}_1={\dot h}_1^{\prime}
+{1\over r}{\dot h}_1-{1\over r}{\dot a}_1+{1\over 2}
({a_0}^{\prime}{\dot a}_0+{b_0}^{\prime}{\dot b}_0+2
{c_0}^{\prime}{\dot c}_0)
\numbereq\name{\eqfcob}
$$
Furthermore, if we introduce $\beta={\dot b}_1-{\dot h}_1$, equation
(\eqfcob) takes the simple form
$$
{\beta}^{\prime}+{3\over r}{\beta}=f_1+f_2
\numbereq\name{\eqzanio}
$$
where $f_1=-{2\over r}{\dot h}_1-{1\over r}{\dot a}_1$ and
$f_2={1\over 2}
({a_0}^{\prime}{\dot a}_0+{b_0}^{\prime}{\dot b}_0+2
{c_0}^{\prime}{\dot c}_0)$
Using now the new set of coordinates $s, \theta$ and integrating (\eqzanio),
we find that
$$
\beta={{{\kappa}G^2M^2}\over {s^2}}({{14}\over 9}-{{56}\over 9}
{\cos^2{\theta}}-{3\over 2}{\cos^4{\theta}})\cos^2\theta
\numbereq\name{\eqtzena}
$$
Inserting this expression into ${\dot b}_1=\beta+{\dot h}_1$
and integrating, we derive
$$
b_1={1\over 3}{{G^2M^2}\over {s^2}}+{{17}\over {12}}{{G^2M^2}\over {s^2}}
{\cos^2{\theta}}.
\numbereq\name{\eqasnj}
$$
We still need to calculate the contribution from the residual
coordinate transformations generated by $v_1, u_1$. According to equation
(\eqrealm), $b_1$ transforms as follows
$$
b_1 \mapsto b_1+b_{0}^{\prime}u_{0}+{\dot b}_{0}v_{0}-2{\kappa}v_1
+2u_{1}^{\prime}-u_{0}^{\prime2}+e^{2\kappa y}v_{0}^{\prime2}
\numbereq\name{\eqzagko}
$$
with $v_1$ and $u_1$ given by equations (\eqhaniwta).
If we integrate the second of these equations, we derive an
expression for $u_1$
$$
u_1={4\over 9}{{G^2M^2}\over {s{\sin^2{\theta}}}}-{1\over 8}
{{G^2M^2}\over s}{{\cot^2{\theta}}\over {\sin{\theta}}}+{\psi}
\numbereq\name{\eqalves}
$$
where $\psi=\psi(r)$ is an arbitrary function of $r$.
Combining equation (\eqtzena) and the expression for
$\delta b_1$, we find $b_1^P$. The Neumann bondary condition
is satisfied for arbitrary $\psi$. Imposing though the requirement
that $b_1^P$ reproduces the standard Schwarzschild metric for
large $r$ determines $\psi$
$$
{\psi}(r)=-{{43}\over {72}}{{G^2M^2}\over r}
\numbereq\name{\eqmiklant}
$$
Combining relations (\eqasnj), (\eqzagko), (\eqhaniwta),
and (\eqmiklant), we find
$$
\eqalign{
b_1^P={{G^2M^2}\over {s^2}}&\Big [{1\over 3}
+{8\over 9}{\sin{\theta}}+{8\over 9}{\csc\theta}+{5\over 3}{\csc^2{\theta}}
-{{16}\over 9}{\csc^3{\theta}}\cr
&+(
{{17}\over {12}}+{2\over 3}{\csc\theta}+{11\over {12}}{\csc^4{\theta}})
{\cos^2{\theta}}-{1\over 9}{\csc^6{\theta}}{\cos^4{\theta}} \Big ]
\cr}
\numbereq\name{\eqviopb}
$$
It follows from the definition of $h_1^P$ as well as the expressions
of $a_1^P$ and $b_1^P$, that
$$
\eqalign{
c_1^P=
{{G^2M^2}\over {s^2}}&\Big [-{1\over 6}
+{4\over 9}{\sin{\theta}}-{7\over 6}{\csc^2{\theta}}
+{8\over 9}{\csc^3{\theta}}+(
-{3\over 4}+{4\over 3}{\sin\theta}-{1\over {9}}{\csc{\theta}}\cr
&-{5\over {12}}{\csc^4{\theta}})
{\cos^2{\theta}}
+({8\over 3}+{\csc\theta}
-{1\over {36}}{\csc^6{\theta}}){\cos^4{\theta}} \Big ]
\cr}
\numbereq\name{\eqvibno}
$$
What is remarkable here is that without any further degrees of freedom,
we have already
$$
c_1^P=O({{G^2M^2}\over {{\kappa^2}r^3}})\simeq 0
\numbereq\name{\eqafondg}
$$
for $\cos\theta={1\over {\kappa}r}$ and $r>>1$. Thus, we
have recovered the standard Schwarzschild metric at large distances
on the {\it physical brane} to the first order of nonlinearity
$$
ds^2\simeq -(1-{{2GM}\over r}+\cdots)dt^2+(1+{{2GM}\over r}
+{{4G^2M^2}\over {r^2}}+\cdots)
dr^2+r^2d\Omega^2
\numbereq\name{\eqagge}
$$
where the dots indicate terms of order $O({{G^3M^3}\over r^3})$ and higher.

\newsection Conclusions.%

In this section we will recapitulate what we have done in this paper. We
considered the most general ansatz for a static, axially symmetric
metric in $D=4+1$ dimensions in the Randall-Sundrum framework
and derived the
Einstein equations. We subsequently  solved the equations
to first order in the gravitational coupling and found an exact,
singularity free solution of the linearized equations.
This solution describes the gravitational field of a spherically
symmetric mass distribution confined on
the {\it physical brane}. At distances far away from the material source and
on the {\it physical brane} this solution coincides with the four dimensional
Schwarzschild metric. Thus
we confirmed that all tests of Linearized General Relativity are
satisfied. Furthermore,
we proceeded to examine the solution of the Einstein
equations to the second order in the gravitational
coupling $GM$. We derived
an approximate expression for the gravitational field which is valid to
second order in the gravitational coupling in a region far away from the source.
When this expression is confined on the {\it physical brane}
it coincides again
with the standard Scwarzschild form to second order and thus
reproduces the correct formula for the precession angle of Mercury. These
results reinenforce the belief that the Randall-Sundrum scenario
reproduces General Relativity on the {\it physical brane} to all
orders in the gravitational coupling. We hope to report soon
our results towards this direction.

\vskip .1in
\noindent
{\bf Acknowledgments.} \vskip .01in \noindent
We would like to thank N. Khuri, J. T. Liu, Y. Pang and M. Porrati
for useful discussions.
This work was supported in part by the Department of Energy under Contract
Number DE-FG02-91ER40651-TASK B.

\noindent
{\bf{Appendix}}

The homogeneous equation (\eqbcud),
$$
\Big[{1\over {s^3}}{{\partial}\over {\partial s}}
(s^3{{\partial}\over {\partial s}})+{1\over {s^2{\sin^2{\theta}}}}
{{\partial}\over {\partial\theta}}({\sin^2{\theta}}{{\partial}\over {\partial\theta}})
-{3\over s}{{\partial}\over {\partial s}}
+{3\over {s^2}}{\tan{\theta}}{{\partial}\over {\partial\theta}}\Big]{\cal F}
(s, \theta)=0
\numbereq\name{\eqreni}
$$
is a Quasi-Laplace since the first two terms inside the bracket is
the four-dimensional Laplacian acting on an axially symmetric function.
Inspired by the large 
$s$ expansion in (\eqapnvh), we look for the solution of the form
$$
{1\over s^n}\sum_{l=0}^{m}F_l(\theta)\ln^l\kappa s
\numbereq\name{\eqrenoi}
$$
with $n$ a positive integer. Substituting it into (\eqreni), we obtain
a system of $m$
ordinary differential equations for $F_l$'s
$$
\Big[{1\over {{\sin^2{\theta}}}}
{{\partial}\over {\partial\theta}}({\sin^2{\theta}}{{\partial}\over {\partial\theta}})
+3{\tan{\theta}}{{\partial}\over {\partial\theta}}+n(n+1)\Big]F_l=
(l+1)[(2n+1)F_{l+1}-(l+2)F_{l+2}]
\numbereq\name{\eqayala}
$$
If we further introduce a new parameter $\varrho=\cos^2{\theta}$, the
differential equations (\eqayala) are transformed into
$$
\Big[{\varrho}(1-\varrho){{d^2}\over {d{\varrho}^2}}+(-1-{1\over 2}{\varrho})
{d\over {d\varrho}}+{1\over 4}n(n+1)\Big ]F_l=
{1\over 4}(l+1)[(2n+1)F_{l+1}-(l+2)F_{l+2}]
\numbereq\name{\eqasioc}
$$
Since $F_{m+1}=F_{m+2}=0$, the equation for $F_m\equiv F$ is a
hypergeometric equation
$$
{\varrho}(1-\varrho){{d^{2}F}\over {d{\varrho}^2}}+(\gamma
-(\alpha+\beta+1){\varrho})
{{dF}\over {d\varrho}}-{\alpha}{\beta}F=0
\numbereq\name{\eqagios}
$$ 
with $\alpha={n\over 2}$,
$\beta=-{1\over 2}(n+1)$ and $\gamma=-1$. The other
equations are of the same 
hypergeometric type but with inhomogeneous terms. The set
of equations can be 
solved successively for $F_l$ in a descending order of the subscript $l$. 
For odd $n$, $\beta$ is a negative integer and there is
a polynomial solution. 
The companion solution, however, is nonanalytic
at $\varrho=0$ or $\varrho=1$ or 
both. The terms of the large $s$ expansion (\eqapnvh)
contains 
odd powers of $s$ and the power of the logarithm $\ln\kappa s$ does appear 
in association with the polynomial solution. For even $n$, the solution can 
be written as 
$$
F={\bar F\over \sqrt{1-\varrho}}
\numbereq\name{\eqroidn}
$$
with $\bar F$ satisfying a different hypergeometric equation with
$\alpha={1\over 2}(n-1)$, $\beta=-{n\over 2}-1$ and $\gamma=-1$. This new 
hypergeometric equation possesses a polynomial solution. But the 
square root in (\eqroidn) does not make it an analytic solution for $F$.
The case with $n=2$ is of special interest. In this case one of the solutions
is
$$
w_1(\varrho)={{\varrho}^2\over {\sqrt{1-\varrho}}}
={\cos^3{\theta}}{\cot{\theta}}.
\numbereq\name{\eqpatsaz}
$$
Given the solution $w_1$ we can derive the second solution by means of
the standard formula, $w_2(\varrho)=w_1{\int^{\varrho}}d{\xi}w_1^{-2}
e^{-{\int^{\xi}}d{\eta}p(\eta)}$ where $p(\eta)
=-{{1+{1\over 2}{\eta}}\over {{\eta}(1-\eta)}}$. Subsequently we find
that the second solution to the homogeneous differential equation is given by
$$
\eqalign{
w_2(\varrho)&=-{1\over 2}-{3\over 4}{\varrho}+{3\over 8}{{\varrho^2}\over
{\sqrt {1-\varrho}}}{\ln {{1-{\sqrt {1-\varrho}}}\over {1+{\sqrt {1-\varrho}}}}}\cr
&=-{1\over 2}-{3\over 4}{\cos^2{\theta}}+{3\over 8}
{\cos^3{\theta}}{\cot{\theta}}
{\ln {{1-\sin{\theta}}\over {1+\sin{\theta}}}}\cr}
\numbereq\name{\eqlembo}
$$

\immediate\closeout1
\bigbreak\bigskip

\line{\twelvebf References. \hfil}
\nobreak\medskip\vskip\parskip

\input refs

\vfil\end

\bye